\documentclass[12pt]{article}

\pdfoutput=1 
\usepackage[applemac]{inputenc}
\usepackage[T1]{fontenc}
\usepackage{amsmath}
\usepackage{amsfonts}
\usepackage[french]{babel}

\usepackage{geometry}                
\usepackage[parfill]{parskip}    
\usepackage{graphicx}
\usepackage{amssymb}
\usepackage{epstopdf}
\usepackage[all]{xy}

\usepackage{amsthm} 
 
\theoremstyle{definition} 
 
\theoremstyle{axiom}

\theoremstyle{remark} 
 
\theoremstyle{plain}

\theoremstyle{plain}

\begin{document}

\title{On von Weizs\"{a}cker's philosophy of Quantum Mechanics}
\author{Daniel Parrochia}
\date{University of Lyon (France)}
\maketitle

\textbf{Abstract.}
We are interested here in the program of reconstruction of quantum mechanics of the German physicist and philosopher Carl Friedrich von Weizs\"{a}cker, which still has some supporters today. In the major part of this article, we limit ourselves to examining purely epistemological and philosophical questions. The theory of the German physicist is often interpreted with reference to Kant, but the most of the time in a rather vague way (categories of understanding, theory of experience). We can afford to be more precise. First we situate the theory of fundamental alternatives (or Ur-alternatives) of von Weizs\"{a}cker in the lineage of the Kantian theory of the transcendental Ideal. We then show that the physicist only substitutes for the classical logic, on which Kant relied, a quantum logic allowing to generate, from the alternatives, via the local isomorphism between certain spinor groups and groups linked to space-time, the whole of physical reality to which we have access. After examining some problems, we finally show how this perspective leads to a quantum theory of information. \\


\section{Introduction}

The German physicist and philosopher Carl Friedrich von Weizs\"{a}cker (1912-2007) is a singularity in the landscape of Germanic physics. Son of diplomat Ernst von Weizsäcker (1882-1951) and elder brother of the former German president Richard von Weizsäcker, he came from an illustrious family who gave Germany priests, theologians, doctors, scientists, politicians. As an elder, he inherited, on the death of his father, the title of "Freiherr" (baron). Between 1929 and 1933, von Weizsäcker studied physics, mathematics and astronomy in Berlin, Göttingen and Leipzig, under the aegis of Werner Heisenberg and Niels Bohr. His thesis supervisor is Friedrich Hund (1896-1997), a quantum physicist close to Bohr, at the origin of the famous “rule of Hund”, an empirical rule useful in spectroscopy to find the fundamental term of an atom. Working in the company of Hans Bethe on the binding energy between nucleons and the nucleus processes ires within stars, von Weizs\"{a}cker discovers a formula describing stellar nuclear processes, since called the Bethe-Weizsäcker formula. He is also, with Bethe, at the origin, in 1937, of the understanding of the cyclic process of nuclear fusion in stars (Bethe-Weizsäcker process (see \cite {Wei1}). From 1938, he develops an accretion theory to explain the formation of the solar system from the contraction of a rotating cloud of gas and dust (see \cite {Wei2}). His work, published between 1944 and 1948, rapidly found a consensus in the scientific community. After the war, during which, like Heisenberg, he remained in Germany and continued his research in atomic physics (including with a view to building nuclear weapons), he became increasingly interested in epistemological and philosophical questions, developing one of the first forms of "quantum logic" and implementing a program of reconstruction of quantum physics. From 1957 to 1969, he held the post of professor of philosophy at the University of Hamburg. During the 1970s, he will create, with the Indian philosopher Gopi Krishna, a foundation "for Western sciences and Eastern wisdom". After his retirement in 1980, he will continue his work of conceptual redefining quantum physics and will lead, at the same time, ethical and political considerations, practically until his death.

\section{General orientation of thought}

Von Weizs\"{a}cker, it seems, has always been a philosopher at heart, and did physics at the age of 30 - on the advice of Heisenberg, by the way - only to practice better. philosophy in his fifties. It is therefore not surprising that the general orientation of his work, including physics, is in fact deeply philosophical. The preferences of the physicist are expressed in favor of what one might call a sort of experimental Platonism, perhaps even a "transcendental" Platonism, as will be seen later. His writings, in any case, manifest multiple references to Plato and even to Parmenides – of which one might think that the direct consequence in physics could to be a definite bias in favor of ideas or theory. Yet this is not the case. A revealing anecdote on this subject is that which Von Weizs\"{a}cker recounts concerning this American physicist\footnote{It's funny enough that von Weizs\"{a}cker prefers to mention this physicist – whose name he doesn't mention anyway – rather than Bohr who, in the BKS theory, actually made the same mistake.} who, in 1935, believed to have experimentally refuted the principle of conservation of energy: in the case of high-energy atomic processes, it was not, according to this searcher, strictly verified and therefore had only a statistical value. Heisenberg, after reading his article, rejected it: "He measures badly" he asserted. However, comments von Weizs\"{a}cker, Heisenberg's rejection was not based on {\it disdain} of experiment, but, on the contrary, on his idea that experiment {\it had to be} a {\it real} experiment, and also, on the correlative observation that the so-called experiment of the physicist in question did not meet the usual requirements. It follows, concludes von Weizs\"{a}cker, that knowing what an experiment should be is not, in general, really clear and requires in fact philosophical and epistemological considerations to be able to be clarified (see \cite {Dri2}, 30). Basically, experiment is necessary, but to experiment is not to understand. And since, unlike Popper, von Weizs\"{a}cker tends to think that no experiment can refute a theory, he reminds us that it is still beauty and simplicity that Heisenberg – rightly or wrongly? – referred to, to judge the relevance of a physical advance\footnote{We know that today, it is precisely such a position that is contested by some physicists (see \cite{Hos}).}.

One will not seek, however, in the philosophical writings of von Weizs\"{a}cker, an approach to past philosophies which is irreproachable from the point of view of the history of philosophy. His references to Greek philosophers or to thinkers closer to us are magnetized by the transcendental perspective in which he places himself. Very influenced by Kant, in fact, von Weizs\"{a}cker essentially sees metaphysics as a meta-physics, in other words, a theory which must state conditions of possibility of physics. But he does not intend, however, to resolve this question in a scientific way or in the manner of logical positivism. Von Weizs\"{a}cker is not only convinced that physics is in history, but that time is, in fact, at the basis of human experience\footnote{Heidegger whom he knew and to whom he refers, certainly also influenced him, especially since the Heideggerian reading of Kant's {\it Critique of Pure Reason}, insists on the importance of time in our experience of the world.}. The co-perception of a difference between sensitive object and the "eidos" – in other words the form, the regularity which accounts for it – make of philosophy a veritable theory of phenomenal appearance where idealist and empiricist positions have never ceased to clash in the course of history (see \cite {Dri2}, 73). Von Weizs\"{a}cker, who does not hesitate to bring together the "$\nu$o\~{u}$\varsigma$ "(or spirit) of the Hellenes and the transdendental subject, attempts to show, in the light of the Greek philosophers (Plato and Parmenides above all), that the quest for physics – including quantum mechanics – is in fact a new research of stable forms and, beyond that, the quest for a true unity of nature. In this context, God is another name for the One and the {\it Timaeus} becomes a kind of model for the physical process, of which we also find equivalents in other cultures (see \cite {Dri2}, 119 ), and this, even if modernity deviates a lot from it.

With some adaptations however – thus the quantum of action can become a kind of new "One" – it is possible to reinterpret modern physics in philosophical terms, finding in Plato the fundamental elements which should be combined, or in Aristotle, an inspiration to found a physical – and no longer mathematical – theory of the continuum. The approach is all the more impressive since von Weizs\"{a}cker is probably one of the last physicists to know ancient Greek and to be able to go into the details of these texts. The result of this research was rather fruitful: the meditation on the hypotheses of {\it Parmenides} or those of {\it Timaeus}, is certainly, as we will see later, one of the origins of the theory of fundamental alternatives (or Ur-alternativen) which must be put, for the physicist, at the foundation of a new theory of physics \footnote {Another source would be this tendency of German thought, since Goethe and his {\it Essay on the metamorphoses of plants}, to seek "Ideal prototypes", "Urformen" or "Urphänomenen", in other words original configurations capable of containing in germ later forms, of defining the organization, and of explaining the development of all that follows. We would find also this tendency in the humanities, for example in the work of Walter Benjamin.}.

\section{The idea of alternative}

We know that in standard logic - say, more precisely, in 2-valued calculus of propositions (where we traditionally represent the value "true" by 1 and the value "false" by 0), the alternative (w) is distinguished from the disjunction ($ \vee $) and can be defined as the negation of equivalence ($ \equiv $). Given two propositions $ p $ and $ q $, we have the following truth tables:

\[
\begin{matrix}
p & q & p \vee q & p\ \textnormal{w}\ q  & p \equiv q \\ \hline
1 & 1 & 1 & 0 & 1 \\
1 & 0 & 1 & 1 & 0\\
0 & 1 & 1 & 1 & 0\\
0 & 0 & 0 & 0 & 1
\end{matrix}
\]

Heisenberg readily regarded symmetry as a fairly natural starting point for a physicist. Even philosophically justified, this reasoning left von Weizs\"{a}cker dissatisfied. This is why, no doubt, he sought to explain to himself the origin of symmetry, and he found it in the existence of fundamental alternatives in our questioning of nature. Such reasoning led him to set up the abstract program of a possible reconstruction of physics in terms of yes-no alternatives, which he later called “Ur-theory”.

The expression – if not the oldest, at least the most concise – of the theory of alternatives\footnote{According to Holger Lyre (see \cite {Lyr}), it seems that the idea dates back to 1954.} such as it was developed by CF von Weizs\"{a}cker, appears in the book that Heisenberg published in 1969, under the title {\it The Party and the Whole} (Der Teil und das Ganze). Heisenberg thus relates the remarks of his former student:

“Any reflection on nature must inevitably take place in large circles or spirals. For we can understand something about nature only on condition of reflecting on it; and all of our behavior, including our thinking, is rooted in the history of nature. In principle, therefore, our reflection could begin at any point. But our thinking is so made that it seems reasonable to start with the simplest; and the simplest is the alternative: yes or no, to be or not to be, good or bad. As long as one reflects on such an alternative as one does in everyday life, there is nothing more to be learned from it. But we know well, thanks to quantum theory, that an alternative does not necessarily require the answers 'yes' or 'no', but that there are still other answers – complementary to those – which consist in fixing a certain probability for the 'yes' or for the 'no', also having a prediction value. The possible responses therefore form a continuum. Mathematically, it is the continuous group of linear transformations of two complex variables. This group already contains the Lorentz group which is associated with the theory of relativity. So, if one asks the question, about any of the possible answers, whether it is right or not, it is already a question about a space which is related to the continuous real-world space-time." (\cite {Hei}, 329-330).

The conversation contains in germ much of the past, present and future work of Von Weizs\"{a}cker. This one assumes in substance that by developing a succession of alternatives, one will be able to join the structure of the group underlying to quantum mechanics, then, from there, reconstruct the space or the space-time of phenomena. How is this possible?

Mathematically, the group of linear transformations of two complex variables mentioned by von Weizs\"{a}cker is GL($ 2, \mathbb{C} $) or, when the variables are normalized to 1, the special linear group SL($ 2 , \mathbb {C} $), of which a useful subgroup in physics is SU(2), or special unitary group of degree 2, which is the group of unit matrices $ g $ with complex coefficients of dimension $ 2 \times 2 $ and determinant 1. We have, explicitly:

	\[
\textnormal{SU}(2) = \left\{g = \begin{pmatrix}
\alpha & -\bar{\beta} \\
\beta & \bar{\alpha}\
\end{pmatrix}
\right\},
\alpha, \beta \in \mathbb{C}, |\alpha|^2 + |\beta|^2 = 1.
\]
		
	It turns out that this group SU(2), which is used to represent the spin of the particles, is homomorphic to the special orthogonal group SO(3) of real three-dimensional space. We know, in fact, that SU(2) is diffeomorphic to the 3-sphere $ \mathbb {S}^3 $ by the following mapping:

\[
\begin{array}{rcl}\varphi :S^{3}\subset \mathbb {R} ^{4}&\to &{\text{SU}}(2)\\ \\

(a,b,c,d)&\mapsto &{\begin{pmatrix}a+\mathrm {i} b&-c+\mathrm {i} d\\c+\mathrm {i} d&a-\mathrm {i} b\end{pmatrix}}.\end{array}
\]

The diffeomorphism $ \varphi $ passes the multiplication of SU(2) to $ \mathbb {S} ^ 3 $, which gives the multiplication of quaternions. SU(2) is therefore isomorphic to the group of unit quaternions. Since the quaternions represent rotations in 3-dimensional space, there exists a surjective homomorphism of Lie groups SU(2) $ \rightarrow $ SO(3) of kernel {+ I, –I}, the following matrices forming a basis of the Lie algebra $\mathfrak {su}$(2):

\[
\mathrm {i} \,\sigma _{x}= \begin{pmatrix}0&\mathrm {i} \\\mathrm {i} &0\end{pmatrix}
\]
\[
 \mathrm {i} \,\sigma _{y}=\begin{pmatrix}0&1\\-1&0\end{pmatrix}
 \]
\[
 \mathrm {i} \,\sigma _{z}=\begin{pmatrix}\mathrm {i} &0\\0&-\mathrm {i} \end{pmatrix}
\]
(where i is the « imaginary unit »)

	These $\sigma$ matrices (called “Pauli matrices”) being used to represent the spin of the particles, which always takes the form of a fundamental alternative, we can therefore consider SU(2) as a typical model of the alternatives in question. Now, the fact that SU(2), the symmetry group of a binary quantum alternative, is locally isomorphic to the three-dimensional rotation group SO(3) in Euclidean space, allows us to conceive that the latter can be, in some way, a consequence of the fundamental alternatives, the connection being made between spinor structures and space-time structures.

This theory devised by von Weizs\"{a}cker in the 1960s, and reported informally by Heisenberg in the passage we have quoted, has fascinating aspects.

It should first be noted that the "circularity" mentioned in the first lines of the text was already mentioned by the author as one of the two possible forms of complementarity from his first article on "Komplementarit\"{a}t und Logik"(see \cite {Wei3} and the commentary on \cite {Jam}, 91-104), a paper to which we will come back later.

Second, the fact that Von Weizs\"{a}cker observes that the alternative is present in everyday life without being able to get much out of it is a seemingly innocuous remark, but in fact fraught with meaning. We must imagine the alternatives as present at all times in the life of an entity capable of decision-making, which makes any decision a particular act which is part of the set of possible decisions responding to the set of possible alternatives under which the universe can appear. In this sense, the behavior of a large organism is just as enigmatic as that of a small particle. In his novel {\it Cosmos}, the Polish writer Witold Gombrowicz made of it an object of meditation, pointing out the trap that such alternatives constitute for human reason – something Buridan had probably already seen or, if not him, the one to whom we owe the story of the donkey hesitating between a meadow and a bucket of water:
	
	"In front of me two small stones, one on the right, one on the left, and a little further on the left the creamy spot of a corner of the earth that the ants had turned over (...) I was going to pass between these two stones , but at the last moment I made a small gap to pass between one of the stones and the small corner of upturned earth, it was a minimal gap, nothing at all ... and yet this very slight gap was unjustified and that, it seems he puzzled me... so, mechanically, I move away again a little bit to pass between the two stones, as I had originally intended, but I experience some difficulty, oh very weak, coming from the fact that, after these two successive gaps, my desire to pass between the stones has now taken the character of a decision, not very important of course, but a decision all the same. What nothing justifies because the perfect neutrality of these things in the grass does not authorize a decision: what difference is it to go through here or there? (...) but the decision, after the few seconds which have just passed, has become more decisive, and how to decide since everything is worth? ... so I stop again. And, furious, I put my foot forward again to pass, as I now want, between the stone and the wedge of the earth, but I see that if I do so after two false starts, it will no longer be a normal walk, but something important ... So I choose the road between the corner of the earth and the root ... but I realize that it would be acting as if I was afraid, so I want to pass between the stone and the wedge of dirt, but the hell, what the fuck is it, I'm not going to stop like that in the middle of the road, what the hell would I be fighting ghosts ? What is it, What is it? (...) I didn't move. I stayed there, standing. This attitude was becoming more and more irresponsible and even insane, I was not allowed to stay there like this, \textsc{impossible}! (...) It was then that I moved. All of a sudden, I tipped all this impossibility into me and went easily without even knowing where, because it didn't matter, and thinking of something else: that the sun was setting earlier here, at because of the mountains. Yes, the sun was already low enough. I walked through the meadow towards the house, whistling, lit a cigarette and only a faint memory of the scene remained, like a vague residue." (\cite{Gom}, 174-176)
	
	The idea that after all, everything could make sense, so that any alternative and any decision that resolves it could have an unsuspected importance, is a familiar situation that one also meets in certain games governed by a strong determinism. In a game like Fan tan, for example, where two players are in the presence of $ p $ pile of matches and take turns choosing a pile (not empty) to remove the number of matches they want (at least one), so that the winner is the one who removes the last match, the situation is entirely determined from the outset. One proves (see \cite{Ber}, 300-309) that a choice is winning if the digital sum of the numbers of matches left in the different piles is zero. In the variant made famous by Alain Robbe-Grillet's film, {\it Last Year at Marienbad}, and where, this time, it is the one who takes the last match who loses, any choice is a winner, obviously, if the digital sum of the numbers of matches left in the different piles is not zero\footnote{The notion of "digital sum" is obtained, from the decimal writing of numbers, by switching to binary writing, then by replacing this writing by what is called the “binary expansion” of the number (we read the binary number backwards, and we separate its components by commas). For example: 3 = 11 = (1, 1); 7 = 111 = (1, 1, 1). Hence 3 + 7 = (1, 1) $ \dot {+} $ (1, 1, 1) = (0, 0, 1) = 100 = 4. This rule follows from the formalization of the game in Graph theory, where the heaps are represented by successive $ G_{i} = (X_{i}, \Gamma_{i}) $ graphs whose vertices $ x_{i} $ represent each time the player's position. For any heap, the method is based on the computation of a Grundy function $ g (x) $ from which we deduce a kernel $ S = \{x \ | \ g (x) = 0 \} $. The gain is assured, in the case of Fan tan, if we choose each time a position $ x = (x_{1}, x_ {2}, ..., x_ {n}) $ for which $ g_{1} (x_{1}) \dot {+} g_{2} (x_{2}) \dot {+} \cdots \dot {+} g_{n} (x_{n}) = 0$.}.

Does quantum physics make us escape this world of deterministic laws? This is one of the challenges of the theory of alternatives, as it will be explained by von Weizs\"{a}cker in various texts. Let us now try to detail the author's progress in the development of his theory.

\section{The quantum logic of von Weizs\"{a}cker}

Although part of the circle of Heisenberg and Bohr, at the origin of the Göttingen-Copenhagen interpretation of quantum mechanics, von Weizs\"{a}cker, from the mid-1950s, sought to develop a logical understanding of this one, which came from a slightly different point of view. As Max Jammer recalls, its first formulation appeared on the occasion of Niels Bohr's seventieth birthday (October 7, 1955) with the first of three articles devoted to question.

 Von Weizs\"{a}cker's logic of  complementarity is constructed as a modification of the logic of contingent propositions, in particular that of simple alternatives (einfache Alternative) mentioned above, as it is presented in experiments such as that of Young's slits, when the particle passes either through slit 1 or through slit 2. Von Weizs\"{a}cker derives the rules of his logic from the very situation which is described by quantum mechanics.

As we know, the state function $ \psi $, behind the diaphragm formed by the slits, is only determined by the two complex numbers $ u $ and $ v $ defined by the equation:	
	\[
	\psi = u \phi_{1} + v \phi_{2}
\]
where $uu^* + vv^* = 1,\ \phi_{1}$ and $\phi_{2}$ propositions characterized by $ (u, v) $ and which differ from $ a_ {1 } $ and of $ a_{2} $ are said to be "complementary to $ a_{1} $ and $ a_{2} $". The complementarity therefore takes on a purely logical character here: if one of the two propositions is true or false, the complementary is neither true nor false.

We find an illustration of this idea in the quantum theory of spin-$\frac{1}{2}$ particles, which seems to have inspired it. In this example, the vector $ (u, v) $ is simply the two-component Pauli spinor. If a given magnetic field is directed in the positive direction $ z $, then the original alternative lies between the two propositions:

$a_{1}$: The spin is in the positive direction $ z $.

$ a_{2}$: The spin is in the negative direction $ -z $.

To any spinor $ (u, v) $ then corresponds a direction of spin orientation which, if $ a_{1} $ has the value $ u $ and $ a_{2} $ the value $ v $, can be defined, in terms of direction cosines by the formulas:
\[
x = uv^* vu^*; \qquad y = i(uv^* vu^*);
\]
\[
z = uu^* - vv^*;
\]\[
x^2 + y^2+z^2 = 1,
\]

and, in terms of polar angle $\theta$ and azimuthal angle $\phi$, by the équations:

\[
u = \textnormal{cos} \frac{\theta}{2} \textnormal{exp}(-\frac{i\phi}{2}), \qquad v = \textnormal{sin}\frac{\theta}{2} \textnormal{exp}(-\frac{i\phi}{2}).
\]

If $ \theta $ is neither 0 nor $ \pi $, then the proposition "The spin has the direction $ (\theta, \phi) $" is complementary to the propositions $ a_{1} $ and $ a_{2} $.

With this assignment of truth values, as von Weizs\"{a}cker was able to show, a higher logical level is introduced, because the original alternative now refers to a property of the physical object and we then wonder about the truth of the alternative or, more precisely, of the answer given to it, which is a meta-question: "what are the truth values of the possible answers?" And it is answered in the complementarity logic by the following assertion: “the possible answers are the normalized vectors ($ u, v $).” The complementarity logic was thus introduced into the object language by means of a metalanguage which applies ordinary 2-valued logic. The difference between classical logic and quantum logic can then be formulated as follows: whereas in classical logic, the proposition $ a_{1} $ is equivalent to the proposition "$ a_{1} $ is true ", this is no longer the case in complementarity logic. Although the truth (resp. falsity) of $ a_{1} $ results from the truth (resp. falsity) of "$ a_{1} $ is true", the converse does not hold. If the proposition "$ a_{1} $ is true" is false, $ a_{1} $ is not necessarily true or false, and this, even though "$ a_{1} $ is true" is definitely false.

In the rest of the article and in those that will follow it a few years later (see \cite {Wei4}; \cite {Wei5}), the relationship between language and metalanguage is deepened, Von Neumann's formalism reinterpreted in terms of logic of complementarity and the more elaborate quantifications linked to a quantum field theory are presented as applications derived from the theory proposed here. Finally, the antiparticles are introduced thanks to a generalization of this quantization procedure by taking, no longer {\it one} complex number but rather {\it a pair} of complex numbers as the truth value.

 At the end of his third article (see \cite {Wei5}, 721), von Weizs\"{a}cker remarks that, as the interaction of fields leads to the introduction of several basic alternatives in the theory, which are nevertheless linked in the same spatial space, it would seem more natural to construct a "primordial field" ({\it Urfeld}) in the sense of Heisenberg's latest work, starting from a basic alternative ({\it Grundalternative}), whose interaction with itself should produce all known fields.This idea will receive much further development (see section 6).

In later years, von Weizs\"{a}cker's logic did not find many supporters, apart from those – like Georg S\"{u}ssmann and Erhard Scheibe – had participated in its elaboration. Niels Bohr was hostile to it (see \cite{Jam}, 379), but it was not known whether he had really examined this attempt closely. It is feared that he did not see the general interest of the prospect.

For us, this seems entirely to lie in the possibility of deducing the whole of the physical world (and perhaps the whole of the world in general) from these fundamental alternatives that quantum mechanics has unveiled, which amounts to making the world – in any case the world as it appears to us via quantum mechanics – a gigantic cascade of alternatives (in principle infinite) which, philosophically, generalizes (even if von Weizs\"{a}cker never expressed it thus) the Kantian theory of the {\it Transcendental Ideal}. We will therefore first recall here the gist of this theory.

  \section{From the Kantian "Transcendental Ideal" to von Weizs\"{a}cker's alternatives}
  
  In the {\it Critique of Pure Reason} (2$^e $ section of Book I of the Transcendental Dialectic) (see \cite {Kan}, 487-495), Kant strives to dismantle step by step the mechanism at the origin of the illusion of which consists the idea of God. The illusion comes from the fact that we attribute a being to what is, in reality, only the idea of a "set of all possibilities", an idea to which we arrive from the observation that, just as every concept is subject to a {\it principle of determinability}, which ensures that, of two contradictory predicates, only one must suit it, likewise, everything is subject, as to its possibility, to a {\it principle of complete determination} "according to
which if all the possible predicates of things be taken together with their contradictory opposites, then one of each pair of
contradictory opposites must belong to it"(\cite {Kan}, 488). And although by this set of all possibilities one thinks of nothing more than the set of all possible predicates, this idea can still be clarified in the sense that it must in fact contain only really possible primitive predicates (in other words, derived predicates such as self-contradictory predicates are excluded). Hence a particular mathematical structure, which we have described elsewhere as being a lattice (see \cite {Par}, 148-162).

Everything is therefore a part of a {\it transcendental substrate} "that contains, as it were, the whole store of material from which all possible
predicates of things must be taken" (see \cite {Kan}, 490) and that is nothing but "the idea of a {\it omnitudo realitatis}". It is therefore a question of an {\it ideal} representation which serves as the basis for the complete determination of all things and which is like "the supreme and complete material condition of the possibility of all that exists -- the condition to which all thought of objects, so far as their content is concerned, has to be traced back" (see \cite {Kan}, 491). For Kant, we pass unduly from this set of all possibilities to the idea of a whole of reality and, from there, to the metaphysical (and religious) idea of a supreme being.

It is then interesting to note the way in which operates according to Kant, and in accordance with classical thought, the logical determination of a concept and, so to speak, the transcendental pseudo-determination of the corresponding object.
	
	\subsection{Logical determination of a concept, disjunctive syllogism and transcendental pseudo-determination}

The logical determination of a concept by reason, writes Kant, "is based upon on a disjunctive syllogism, in which the major premiss
contains a logical division (the division of the sphere of a universal concept), the minor premiss limiting this sphere to a certain part, and the conclusion determining the concept by
means of this part" (see \cite{Kan}, 491).

We have already commented (see \cite {Par}, 117-128) on the texts of Kant's {\it Logic} (see \cite {Kan2}) where the philosopher explains what is the "logical division" of the {\it sphere} of a concept or, as we would say today, its {\it extension}. Let us only recall here that, according to Kant – and classical logic – the members of the division must be separated from each other by contradictory opposition, and not by simple contrariety, which means that the division is always dichotomous, because the dichotomy – unlike polytomy – does not require reference to an intuition ({\it a priori} or sensuous) and is based only on the principle of contradiction. Hence this apparently innocuous note from the {\it Critique of Pure Reason} where Kant specifies that "the determinability of every concept is subordinate to the universality ({\it universalitas}) of the principle of excluded middle"(see \cite {Kan}, 488n).

Let us then explain what a "disjunctive syllogism" is, in the sense of classical logic. It is, as Kant reminds us, a valid reasoning where the major is a disjunctive proposition, the minor eliminates one of the members of the disjunction and the conclusion determines the other. Consider the simple example:

(1) The door is open or (exclusive) the door is closed;

(2) The door is not open;

(3) So the door is closed.

It will be noted that the "open" and "closed" predicates are considered here as contradictorily opposed predicates in the sense of classical logic, that is to say that they exclude, as Kant wants it, any "middle", ie any intermediate situation (door ajar, half open, three-quarter open, etc.).

Under these conditions, the succession of more and more precise logical determinations of the concepts starting from the whole of possibility takes place by a cascade of reasoning of the type: $ A $ or $ B $, not $ B $, $ A $.
	
	The transcendental pseudo-determination will proceed in the same way: obviously, as Kant remarks, the concept of a reality in general cannot be divided {\it a priori} since, without experience, we do not know specific kinds of realities understood under gender: this is why we speak of "pseudo-determination". In this context, however, the major of the syllogism cannot be anything but the representation of the whole of reality, which therefore becomes the prototype ({\it prototypon}) from which all things are supposed to derive as more or less defective copies. But what is hidden behind this set is not an original being or the Being of all beings, it is only the transcendental ideal, that is to say the set of all the primitive and not self-contradictory predicates that we have transformed, illusorously, into supreme reality. As the philosopher Gilles Deleuze saw very well (see \cite {Del}, 342), Kant thus reduces God – in this text – to the illusion built on the principle of the disjunctive syllogism.

We will have noted that, the Kantian disjunctions being founded on logical conceptual divisions opposing contradictory predicates, they are in reality alternatives (this, or that, but not both at the same time). But these are alternatives as conceived by classical logic, that is to say, if we pass from predicates to the predicative propositions which are associated with them, propositions which take their truth values in a 2-valued logic and which follow the rules of it.

On the contrary, von Weitzs\"{a}cker's logic is obviously – as we saw above, following Max Jammer's commentary – a particular case of infinite-valued logic\footnote{As it is known, the origin of infinite-valued logics (see \cite{Mal}) lies in the generalization operated by \L ukaciewicz, in 1922, of trivalent logic to a family of many-valued matrices with finite and even infinite number of values. In order to define them rigorously, let us introduce a logical language, with variables, parentheses, definitions of well-formed expressions, as well as a set of connectors of the type $ \neg, \supset, \vee, \wedge, \equiv $. We will call \L ukasiewicz's $ n $-valued matrix (with $ n \in \mathbb {N}, n> 2 $, where $ n = \aleph_ {0} $, or $ n = \aleph_{1} $), a matrix such that the set of truth values is:
\[
 \{0, 1/n-1, 2/n-1,..., 1\},  \qquad \ \qquad \quad \textnormal{if} \ n \in \mathbb{N}, n>2 \ (finite\ case)
\]
\[
\{s/w : 0 ≤ s ≤ w;\ s,w \in \mathbb{N}, w \neq 0\}, \qquad \qquad \textnormal{if} \ n = \aleph_{0}   \ (denumerable\ infinite)
\]
\[
[0,1],			 \qquad \qquad\qquad\ \ \quad\qquad \qquad \quad  \textnormal{if} \ n = \aleph_{1}  \ (continuum)
\]
In such logic, connectors are defined as follows:
\[
\neg  x =: 1 - x;
\]
\[
x \wedge y =: \textnormal{min} (1, 1 - x +y);
\]
\[
x \vee y =: (x  \supset y) \supset y = \textnormal{max} (x,y)
\]
\[
x \wedge y =: \neg(\neg x \vee \neg y) = \textnormal{min} (x,y)
\]
\[
x \equiv y =: (x \supset  y) \wedge (y \supset  x) = 1 - |y - x|
\]
One easily deduces from this last expression the formula for $ x\ \textnormal{w}\ y = \neg (x \equiv y)$, since it is the formula which is complementary to the previous one, ie:
\[
x\ \textnormal{w}\ y =: |y - x|.
\]

In an infinite-valued logic, we can therefore apply the disjunctive syllogism without any problem. Whatever the values of $ x $, of $ y $ and of $ x $ w $ y $, we can always deny an alternative disjunction a part (resp. finite or infinite) to obtain, in fine, its complement (resp. infinite or finite).

If now we replace the values $ x $ and $ y $ by complex numbers $ z $ and $ z '$ identified respectively with the couples $ (u, v) $ and $ (u', v ') $, we can define a value of the alternative in the sense of von Weizs\"{a}cker by the complex number $ | z - z | $, which makes its logic a variant of the logics of \L ukasiewicz. As we will see, it is such that the laws of quantum physics sometimes make it escape ordinary logic.}.

 Transforming in principle any alternative into an infinite alternative, it will singularly complicate the disjunctive syllogism at the origin of the complete determination of concepts or properties. We are going to see what an alternative looks like in an infinite-valid logic of the type of that of von Weizs\"{a}cker and what becomes, in this logic, the disjunctive syllogism. But first, let's look at modern questionings of the Kantian structure.

	\subsection{Modern criticisms of Kant}

	As we have seen, the Kantian {\it transcendental ideal} – as a set of possible predicates or a set of all possibilities – is a set of two-by-two contradictory predicates that can also be represented in the form of a collection of alternatives, since, each time, one of these predicates or contradictory properties can be attributed to a subject, and therefore be the object of a predicative proposition. The one who knows everything would therefore have mastery of the disjunctive syllogism, that is to say, starting from the set of all possibilities, could without a doubt exclude from any subject whatever, the predicates which do not suit it and generally assign the right predicates to the right subjects.

The weakness of contemporary philosophies (French Theory) is often due to the fact that their questioning of classical thought does not lead to solid theories but to unorganised universes where anything is possible – which does obviously make no sense from a physical point of view.

The case of Gilles Deleuze's philosophy is a remarkable example. In his critique of Kantian philosophy, he had noticed that this negative and limiting use of exclusive disjunction (otherwise called alternative) could be opposed to another form of disjunction – inclusive and affirmative. The set of possible predicates is then no longer limited by the negations of successive disjunctive syllogisms, giving rise – if, as in the case described by Kant, one abusively attributes a being to it – to an entirely different figure than that of the supreme being.

The model is found in some writings of Pierre Klossowski. In this latter theory, God, as Being of beings, is replaced by Baphomet, "prince of all modifications" or "modification of all modifications" (see \cite{Del}, 344).
	
	This figure of the "devil" or of the "Antichrist" is then no longer the one which excludes the predicates of a thing by determining it, that is to say, by itself denying the predicates which are not suitable, but the one which opens everything to the infinity of possible predicates by affirming them all.

The Baphomet itself, in traditional iconography, is presented as a bearded figure, with the head of a goat and possessing breasts. In other words, he is at the same time a man, a woman, an animal, etc. \footnote {The famous {\it Dogma and ritual of High Magic of Eliphas Lévi}, gives him the form of a goat standing on his feet. hind legs and bearing horns. His right hand points to, "above the white moon of Chesed", while the other shows "below the black moon of Geburah" and one raised arm bears the inscription "Solve", the other "Coagula". The author underlines his androgyny by adding: "One of his arms is feminine, the other masculine" (see \cite{Dec}, 8-11). With Pierre Klossowki, the effects of Baphomet were mainly limited to mixing the souls of the monks and pushing the laws of hospitality a little further. We stayed in the humor. Taken seriously and carried to its most extreme consequences, the possibility of an infinite combinatorial of predicates leaves a little perplexed: this is the world of surrealism and "transgender".}. Contradictory predicates are allotted to it together.

In Klossowski's fantastic novel where the "souls" or "breaths" of the dead, who can reincarnate, ultimately lack the body to do so and spin around indefinitely, they become entangled, and entangled so that the identities of each are no longer assignable: "also each of these breaths was lost in the scent of another, according to this immediate assimilation with which each was endowed, the most voluptuous as the most chaste, the most criminal as the most innocent" (see \cite{Klo}).
	
	In this world without names – and therefore without identities – also lacks what the virtue of names usually implies (identity usually goes, logically, with non-contradiction, the third excluded, and, morally, with responsibility, guilt, etc.). In other words, where subjects and predicates are entangled, any predicate can qualify any subject.
Instead of the kantian transcendental ideal, we therefore have here what we could call a {\it trancendental infernal}, from which we no longer derive, by paralogism, the idea of God, but the symmetrical illusion: that of Baphomet. Yet this one is it with difficulty – Deleuze does not seem to have seen the problem – the "master" of the new disjunctive syllogism. Indeed, an (infinite) set of predicates being posed, with the rule that any predicate can be associated with any other without any restriction, the major stating this combinatorial can hardly be accompanied by a minor other than itself, except to think that one can subtract an infinity from an infinity. Otherwise, the so-called syllogism boils down to pure tautology.

But we must also observe this: the discrete set of predicates that can be put in bijection with the set $ \mathbb {N} $ of natural integers having the power of the countable ($ \aleph_{0} $), the set of all possible combinations of sequences of predicates is identified with the set $ \mathbb {N} ^ {\mathbb {N}} $, in other words, a set possessing the power of the continuum ($ \aleph_ { 1} $) \footnote {The set $ \mathbb {N}^{\mathbb {N}} $ is the set of sequences of natural integers. By assigning a number to each predicate, then by supposing, as it has been postulated, that we can form all the possible sequences of predicates, we are led to attribute the power of the continuum to this combinatorial.}.

In other words, if we follow the mathematician linguist Solomon Marcus, it has exactly the same power as poetic language (see \cite{Mar1}, 53-54; \cite{Mar2}). You can certainly see the world this way when you are an artist or a writer, but this is surely not how physical reality was formed, which, despite its fantasies, involves much more rigid constraints \footnote {Quantum physics has certainly broadened the notion of “real” by notably introducing new aspects into it (superposition of states, irreducible indeterminacy, non-locality, etc.). In our view, however, this extension does not go so far as to abolish the distinction between real and possible (which is also that of mathematics and physics) and even less the distinction between possible and impossible (which is that of understanding and imagination). The abolition of the distinction between real and possible amounts to maintaining that everything that is mathematical is real (thesis of Max Tergmark). As for the abolition of the distinction between possible and impossible, it amounts to maintaining that everything that one can imagine exists in one and the same sense (thesis of the univocity of being, taken up by Deleuze with its unlimited combinatorial of predicates). One can easily refute these two theses: indeed, physics uses only a very small part of the mathematical structures; on the other hand, anyone who knows mathematics also knows that there are objective impossibilities (for example, theorems of non-existence which, even presented in a positive way – which is always possible – nevertheless translate what one might call facts of "solidarity" or "rigidity": for example, there are no triangles whose side lengths are equal and whose angles are unequal; or again (Liouville's theorem): there are no bounded integer functions which are not constant.}.	

\subsection{The quantum transcendental ideal}

In the case of a transcendental reflection like that of von Weizs\"{a}cker\footnote {The program for the reconstruction of physics is indeed a" transcendental "program, as Gernot Böhme indicates (see \cite{Boh}), but it is less the categories of understanding and the sensuous intuitions that are their source than the fundamental alternatives which refer to the Kantian ideal of pure Reason.}, inspired by quantum experience and the logic which accounts for this, the modification of the Kantian transcendental ideal does not go as far as in the Klossowskian-Deleuzian case: not everything is compatible with everything. The weizs\"{a}ckerian transcendental ideal, if we can thus be expressed, no longer rests, no doubt, on classical oppositions, but it is based on authentic complementarities, quantum complementarities. Intensely, given a subject, not only one or the other of two contradictory properties can be attributed to it, but any proportion of one or the other, without exclusion. At the extensional level, quantum logic is substituted for the computation of classical propositions expressed in 2-valued logic. The ideal is therefore structured, we will see how.

\subsubsection{The finite case}

By placing ourselves in a finite case, we can already see very well the formal relationship existing between the set of possible objects associated with the Kantian transcendental ideal and the set of possible states of a classical physical system. We will see by the same the formal difference existing between these two and the set of possible states of a quantum physical system.

Let 4 (2 by 2) contradictory predicates be denoted respectively by $ p $ and $ p '$, $ q $ and $ q' $. If we ask what are the non-contradictory objects that we can form by combining these predicates by operations of conjunctions or disjunctions, we see that they appear in a distributive and complemented (therefore Boolean) lattice whose upper bound is the set of possible predicates and the lower bound the empty set (see Fig. 1.a)\footnote {In an old book (see \cite{Par}, 153), considering the set of non-contradictory objects {\it stricto sensu}, we eliminated the upper and lower bound of the lattice and reduced it to a matroid: we did not have, then, to compare this structure to a physical system. But it is basically, except for this difference, the same structure.}.

Let now be a physical system, that is to say a part of reality conceived as existing outside the physicist, and sufficiently isolated to be able to be considered in itself. We call "propositions" the well-formed sentences describing the states of this system in response to yes/no questions. On this set, we define a preorder relation $ <$ such that:	
\[
a < b \quad \textnormal{will mean}  \quad a\ \textnormal{true} \Rightarrow b\ \textnormal{true}.
\]
	
	Under these hypotheses, one shows (see \cite {Pir}, 3) that the set of propositions of a physical system (answering questions having the form of alternatives formulated about it) is a complete lattice. The only one difference between the classical situation and the quantum one is that, in the first case, the lattice is distributive, which is not true in the second. More precisely, in classical physics, "$ a \wedge b $ true" always results in "$ a $ true" and "$ b $ true", and $ a \vee b $ true "also implies, by analogy with the logical, “$ a $ true” or “$ b $ true.” These relations in turn automatically lead to the following distributive laws:	
	
	\begin{equation}
(a \wedge (b \vee c) = (a \wedge b) \vee (a \wedge c),
\end{equation}
\begin{equation}
a \vee(b \wedge c) = (a \vee b) \wedge (a \vee c).
\end{equation}
	
	But if (1) follows from the definitions, (2) is not always true – especially in quantum physics.

Given a simple classical physical system, with 4 two by two contradictory propositions $ (a, a ', b, b') $ and such that $ a <b $ hence $ b '<a' $, the system is described by a set of non-contradictory propositions of the type $ a, a' \wedge b, b', a \vee (b \wedge a') = b $, etc. The associated distributive and complemented lattice (Fig. 1.b) has the same structure as that of Fig. 1.a. The upper bound of the lattice is the set of possible affirmations (trivial proposition), the lower bound its opposite (contradictory proposition).

If we now take the example of spin-1/2 particles, again by considering 2 systems $ S $ and $ T $ not co-measurable, we will have, on the one hand, the propositions $ a_{1}, a_{2} , a_{1} \vee a_ {2} = 1, a_{1} \wedge a_{2} = $ 0 and, on the other hand, the propositions $ b_ {1}, b_ {2}, b_ {1 } \vee b_ {2} = 1, b_ {1} \wedge b_ {2} = $ 0. By identifying (respectively) the 2 upper bounds and the 2 lower bounds, we can then represent the system on a single lattice with the shape of a "Chinese lantern" (see Fig. 1.c).
	 
	 \begin{figure}[h] 
	   \centering
	      \vspace{-3\baselineskip}
	   \includegraphics[width=6in]{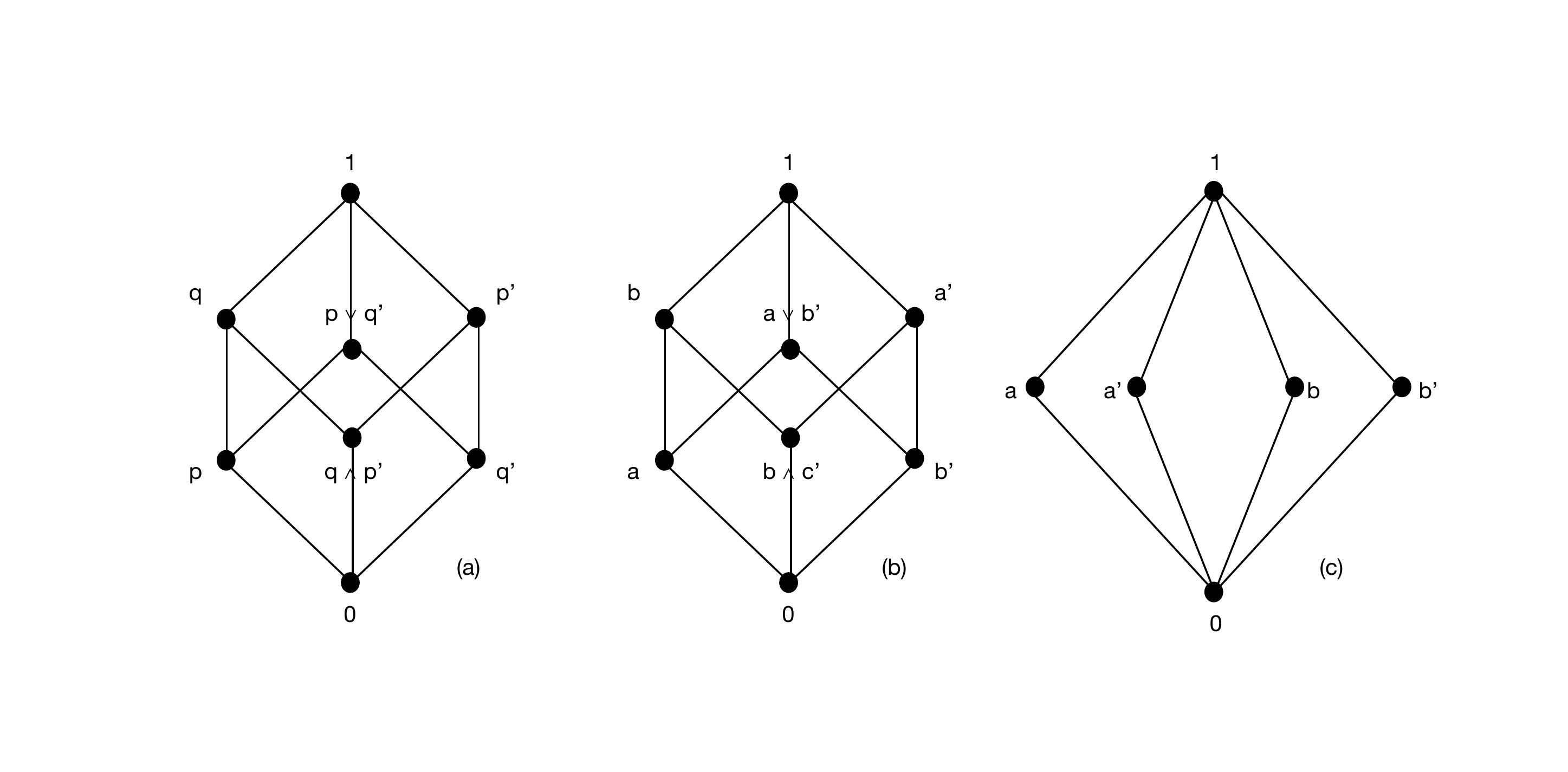} 
	      \vspace{-3\baselineskip}
	   \caption{Kantian lattice(a), classical physics lattice (b) and quantum lattice (c)}
	   \label{fig: Lattice}
	\end{figure}
	
	In this quantum lattice, the modular law $ (a \vee b) \wedge c = a \vee (b \wedge c) $ is always satisfied. So we can call this lattice, as Svozil does (see \cite {Svo}, 27-28), the lattice $ MO_{2} $ ($ M $ for "modular", $ O $ for "orthocomplemented", the index 2 indicating that it is the union of two Boolean algebras $ L (a) $ and $ L (b) $. We can thus note it:
	\[
MO_{2} = L(a) \oplus L(b)
\]
a structure which can be generalized into:
\[
MO_{n} = \oplus_{i=1}^n L(x^j),
\]
if we consider a finite number $ n $ of measures of different spin directions. The resulting structure is the sum of $ n $ classical Boolean algebras $ L (x^j) $ where $ x^j $ indicates the measure of the direction of the spin.

\subsubsection {The infinite case}

The passage to infinity obliges to consider, as we said above, all the possible quantum states, that is to say to give oneself a separable Hilbert space $ H $ (vector space on the set of complex numbers), with infinite dimension and countable base. This space is the analogue of the set of Kant's all possible predicates, when these predicates relate to physical realities.

The set of observables corresponding to possible Kant objects will then be formed by the set of subspaces of Hilbert space, and these constitute an orthocomplemented lattice which, according to Constantin Piron, is not, in general, modular – contrary to what Birkhoff and von Neumann thought –, but only "weakly modular" (see \cite{Jam}, 393).

Finally, the quantum transposition of the disjunctive syllogism supposes a reasoning which can be considered as the transposition, in modern calculus of propositions, of the disjunctive syllogism of classical logic, and which is of the type:

	(1) $p$ w $q$,
	
	(2) $\neg q$,		  \qquad \qquad   (*)
	
	(3) $p$.
	
	However, Heelan (see \cite{Hee}) was able to show that this inference scheme is precisely not always valid in quantum mechanics. Consider, for example, the following case:

$ p $: The electron has an upward spin;

$ q $: The electron has a spin oriented horizontally to the left;

$ p $ w $ q $ being true, if $ \neg q $ is also true, according to classical logic, we should be able to apply the previous scheme (*) and deduce $ p $. However, the quantum situation rather leads to the conclusion that, as soon as $ \neg q $ is true, the spin of the electron is oriented horizontally to the right, and not upwards. The determination of objects from the set of quantum states cannot therefore follow the rules of the disjunctive syllogism.

In other words, not only an infinite number of “complementary” alternatives are juxtaposed with the typical oppositions of classical logic\footnote {Note that, even in the Kantian case where we are only dealing with binary alternatives, the structure of the transcendental ideal is not simple because the division of concepts goes to infinity. Moreover, this division is not unequivocal. A concept can be co-divided in several ways, and this co-division, according to Kant, also goes to infinity.} but, depending on the selection of alternatives that we have practiced, the disjunctive syllogism {\it stricto sensu} can be made inapplicable.

It is quite easy to see that the introduction of a many-valued logic (including infinite-valued logic) will not change anything in the matter, any more than that of an infinitary language (for example, in the sense of Bell and Slomson (see \cite {Bel})). At the concrete level, it is the measurement, causing the "collapse" of the wave function, which will ultimately determine the object as that portion of the possible corresponding to the real. On a purely logical level, the quantum transcendental ideal will therefore not, or not always, lead, by the application of disjunctive reasoning, to the complete determination of a quantum object. In this sense, there is no “master” of the quantum disjunctive syllogism and, at the same time, neither God nor Devil who, by extrapolation, can correspond to it.

It does not seem that von Weizs\"{a}cker perceived this aspect of the question. In any case, he will attempt to clarify the transcendental foundation of physical experiment by placing himself in the finite, which in fact amounts to limit the number of pure fundamental states.	
	
\section{quantum theory «revisited»}

	It is in a text from 1971, published in English in 1980 (\cite {Wei7}, then taken up later in the collection of von Weitzs\"{a}cker's major texts in physics (see \cite {Dri}, 75-109) that the author will develop his true conception of quantum physics – and even of physics at all. For him, indeed, quantum mechanics must achieve the unification of physics by basing it on a particular logic (namely the logic of circumstantial propositions) and by constructing it axiomatically as a general theory for the prediction of empirically decidable alternatives. This construction is finitist\footnote {Obviously, the proponents of a holistic perspective will dispute this point, arguing that there are possibilities which cannot be reduced to ultimate propositions (see \cite {Kha}). But we think that von Weitzs\"{a}cker at least partially escapes this criticism because the number of fundamental elements (Urs), of elsewhere in growth, is not really fixed, if not by an order of magnitude.} in the sense that only a finite number of alternatives can be decided, which supposes, mathematically, the use of Hilbert spaces of finite dimension. This formalism allows to give a unified structure to all physics, leading to a cosmology, a theory of elementary objects and a connection between the two.

Von Weitzs\"{a}cker starts from Copenhagen's interpretation which he believes is correct, but has not been sufficiently clarified. As the whole of physics can be reconstructed from quantum mechanics, it is important to clarify its foundations, and von Weitzs\"{a}cker, borrowing these considerations from one of his students, Michael Drieschner, proposes to first pose the following postulates:	

	A. {\it Postulate of alternatives}: Physics formulates probabilistic predictions concerning the outcome of decisions concerning empirically decidable alternatives\footnote {We can consider that Jauch and Piron had already developed such an idea (see \cite{Jau} and \cite{Pir}).}. An "alternative", in the sense of von Weizs\"{a}cker, can be defined as a complete list of mutually exclusive circumstantial propositions. They are propositions of the type "it's raining in Hamburg" or "this particle is in the position $ x $", in other words, contingent propositions, which can moreover not only relate to the present, but to the past or to the future. If an alternative includes only two, we will then speak of {\it simple alternative}.

B. {\it Postulate of objects}: Answers to an alternative attribute contingent properties to an object. Logically, this means that the responses to an alternative can be formulated as "categorical judgments", which assign a predicate to a subject, called here "object".

C. {\it Postulate of ultimate propositions}: For any object, there exist ultimate alternatives whose answers are ultimate propositions. The particular properties corresponding to the ultimate propositions will be called "states" of the object. For a classical object, a point in phase space represents a state, and the set of all these points represents its ultimate (unique) alternative. For a quantum theoretical object, a one-dimensional subspace of Hilbert space represents a state and any complete orthonormal system represents an ultimate alternative.

	D. {\it Postulate of finitism}: The number of answers to any alternative for a given object does not exceed a fixed positive integer $ n $ which is characteristic of this object. This is the postulate by which Drieschner's approach differs from usual quantum theory. Of course, one can expect that, by choosing $ n $ large enough, it will be possible to avoid contradictions with the predictions of usual quantum theory.

E. {\it Postulate of the composition of objects}: Two objects define a composite object of which they are parts. The direct product of any two ultimate alternatives concerning these two parts is an ultimate alternative of the composite object. If $ a_ {j}\ (j = 1 ... m) $ and $ b_ {k}\ (k = 1 ... n) $ are the answers to the two alternatives, $ a_ {j} \wedge b_ {k} $ (where $ \wedge $ means "and") are the answers to their direct product. This assumption seems quite natural; it has considerable consequences in quantum theory.

F. {\it Postulate of the probability function}: Between two states $ a $ and $ b $ of the same object, a probability function $ p (a, b) $ is defined, giving the probability of finding $ b $ if $ a $, that is, provided that $ a $ is necessary.
	
	G. {\it Postulate of objectivity}: If a certain object really exists, an ultimate proposition about it is always necessary. "A certain object" is an object for which we know how to decide the alternatives, or at least some of the alternatives which characterize it. Obviously, if such an object does not exist at all, then none of its states can be found and, therefore, no ultimate proposition about it can be considered necessary. Now suppose that the object actually exists. This affirmation of existence is a proposition concerning the object. One assumes that any existing object admits alternatives with $ k $ components, with $ k> 1 $. The proposition that the object exists is not an ultimate proposition, for it is implied by any answer to any of these alternatives. Those which relate to composite objects that are not the product of pure states of their component elements, and are considered to have only a virtual existence, will therefore not be accepted as propositions on an object (this makes it possible to exclude the EPR paradox). It will further be assumed that a theory capable of making predictions about the behavior of the object in question is possible; otherwise, the object would not fall under our concept of experiment. From the ultimate propositions and the usual logical connectors, we can then reconstruct the orthocomplemented lattices\footnote {An orthocomplementation on a bounded lattice is a function which associates each element $ a $ with an "orthocomplement" $ a ^ \perp $ of such so that the following axioms are satisfied:	
	
	Complementarity law: $a^\perp \vee a = 1$ et $a\perp \wedge a = 0$. 
	
	Involution Law:  $a^{\perp\perp} = a$. 
	
	Order inversion law: if $ a \leq b $ then $ b^\perp \leq a^\perp $.} of propositions which are those of the von Neumann model of quantum mechanics.

To these postulates, of a rather philosophical nature, will then be added two more precise and more calculating ones:

H. {\it Postulate of measurability}: Any linear operator in vector space which leaves the measure invariant is an observable.

I. {\it Postulate of continuity}: The possible changes of state over time are described by one-to-one mappings, continuous over time, of all the states of the object on itself.

The new axiomatics of Quantum Mechanics then takes the form of a series of rather reasonable axioms:

1. {\it Axiom of equivalence}: If $ a $ and $ b $ are ultimate propositions, $ p (a, b) = 1 $ is equivalent to $ a = b $.

2. {\it Axiom of finite alternatives}: If $ n $ ultimate propositions $ a_ {i} \ (i = 1, ..., n) $, mutually exclusive, are given, then for any ultimate proposition $ b $ :
\[
\sum_{j = 1}^n p (b, a_ {i}) = 1,
\]
 which is a probabilistic version of postulate A.

3. {\it Decision axiom}: For any $ A $, there exists an alternative $ a_ {1} ... a_ {n} $ such that $ a_ {1}
... a_ {e} $ are elements of $ A $ while $ a_ {e + 1} ... a_ {n} $ are elements of $ \bar {A} $. It was assumed in Postulate A that all contingent propositions are decidable. It is now assumed that there are always some ultimate alternatives fitting any decision between $ A $
and $\bar{A} $.

4.a. {\it First axiom of completeness}: For any set of $ k <n $ mutually exclusive ultimate propositions
$ a_ {1} ... a_ {k} $, there exists an ultimate proposition $ a $ with $ p (a, a_ {i}) = 0 \ (i = 1 ... k)$.

4.b. {\it Second axiom of completeness}: For any set of $ n - 2 $ ultimate propositions $ a_ {3} ... a_ {n} $ and any ultimate proposition $ b $, there exists an ultimate proposition $ a_ {2 } $ which excludes all $ a_ {i} \ (i = 3 ... n) $ and $ b $; that is to say, we have:
\[
 p (a_{2}, a_{i}) = 0 \quad \textnormal{and} \quad p (a_{2}, b) = 0. 
\]
	
5. {\it Axiom of indeterminacy}: For two mutually exclusive ultimate propositions $ a_ {1} $ and $ a_ {2} $, there exists an ultimate proposition $ b $ such that $ p (b, a_ {1} ) \neq 0 $ and $ p (b, a_ {2}) \neq 0 $.

6. {\it Axiom of exclusion}: For ultimate propositions, $ p (x, y) = 0 $ implies $ p (y,
x) = $ 0. This is related to the law of double negation $ \bar {\bar {A}} = A. $

Using these axioms, Drieschner – explains von Weizs\"{a}cker – was able to show that:

(a) The set of propositions is a complete non-Boolean lattice.

(b) This lattice is a projective geometry with $ n - 1 $ dimensions.

(c) This lattice is isomorphic to that of the subspaces of a $ n$-dimensional vector space.

(d) In this space, the probabilities define a metric.

After having thus reconstructed quantum mechanics, von Weizs\"{a}cker attempts, in a second part of the article, to found the unity of physics, which leads him to introduce three new postulates:

J. {\it Postulate of approximately stationary cosmology}: The universe can be treated approximately as an object\footnote {This is precisely what a philosopher like Gaston Bachelard contested, probably wrongly.}. Treating the universe as an object is exactly what cosmological models do. The postulate will therefore serve a similar purpose in current theory by allowing suitable terminology.

K. {\it Postulate of ultimate objects}: All objects are made up of binary ultimate objects. In German, von Weizs\"{a}cker called them {\it Urobjekte}, and their alternatives {\it Ur-alternativen}; by fancy, von Weizs\"{a}cker declares to have proposed the abbreviation "Ur "("ur "without capital letter in translation) for this kind of object. The author recognizes that this postulate is trivial as long as one does not specify the law of interaction for the ultimate objects. A $ n $ dimensional Hilbert space can always be described as a subspace of the tensor product of at least $ r $ 2-dimensional Hilbert spaces, where $ 2^{r-1} <n <2 ^ r $ ; the unused dimensions $ 2^{r - n} $ can then be excluded by imposing an interaction law which leads to a rule of super-selection between the two subspaces.

L. { \it Interaction postulate}: The theory of the interaction of ultimate objects (or “urs”) is invariant in the same group as the theory of free ultimate objects. This is a strong and non-trivial hypothesis, which requires, above all, to study free urs.

An Ur reduced to itself is an object in a two-dimensional Hilbert space. It admits the SU(2) transformation group. The physical meaning of this primitive mathematical statement is as follows: The "theory of a free ur" describes not only the variety of states of an ur, but also its law of motion. The equation of motion must be invariant under SU(2). It is easy to find the solutions of such an equation; the state vector can only have a time factor $ e^{- i \omega t} $ common with the independent state $ \omega $, hence the fact that the states themselves, being one-dimensional subspaces, remain unchanged over time. Yet the question is how we know that this is the correct condition to impose on the law of motion for the free ur. This is the form in which the often asked question now arises: why should we postulate symmetry for state space? Von Weizs\"{a}cker formulate the principle involved here in the form of a postulate:

M. {\it Postulate of symmetry}: None of the states of a single ultimate object is objectively distinguishable from the others. An "objective" distinction is here a distinction "by law of nature", as opposed to a "contingent" distinction such as, for example, "the state in which this object currently is"\footnote {Such a postulate is that of any scientific approach aiming at a certain universality. For an equivalent in classification theory, see \cite {Par1}, 29.}; the law of motion should therefore not make a distinction between states.

In this context, the Hilbert space of an ur reduced to itself is a 2-dimensional representation of the group SU(2). The postulate L implies that the Hilbert space of several urs must be a higher dimensional representation space of the same group. Now this group is isomorphic to SO(3), the special orthogonal group of three-dimensional space. The postulates set out above therefore force us to suppose that positional space (i.e. what we usually call “space” or “cosmic space”) is a real three-dimensional spherical space, which seems to suggest that the three-dimensionality of space is a consequence of the quantum theory of ultimate objects.

Now it is clear that the "elementary particles" must be made up of these ultimate objects (or urs). These urs could be, in a way, the elements of a {\it Urfeld} in the sense that Heisenberg envisioned it. The theory having a cosmic dimension, von Weizs\"{a}cker does not hesitate to calculate the total number of urs in the universe, which he deduces as follows: if we assume a radius of the world $ R = 10 ^ {40} $ nuclear units of length $ L = 10^{- 13}$ cm (Planck length), that is $ R = 10 ^ {40} \times 10 ^ {- 13}$ cm, a particle of nuclear moment or a localizable particle in a nucleus should be associated with about $ 10^{40}$ urs. The total number of urs in the universe could thus tentatively be identified with the number of bits (or, as one will say later, of Qu-bits) of information possible in the universe. Assuming, for the sake of simplicity, that there is only one kind of elementary particle, say a nucleon falling under the Fermi statistic, one would estimate that there are as many bits in the world as there are nuclear-sized cells, each of which can be occupied or empty. Under this assumption, this number would be $ N = R ^ 3 = 10^{120} $. The number of dimensions of the Hilbert's space of the universe could then reach $ n_{u} = 2^N $. If a nucleon is made up of $ R $ urs, there should therefore be $ R^2 = 10^{80} $ nucleons in the world, which obviously remains, given the coarseness of the calculation, a simple approximation.

Since elementary particles essentially depend on groups of symmetries, it remains to explain in detail how particles are constructed from urs. But this would suppose the introduction of a new model because the group most generally accepted in field theory being the Lorentz group, we must pass from the present theory, which is not Lorentz-invariant, to one which is. This refers to the approximate nature of postulate J, which will need to be revised. To this end, von Weizs\"{a}cker then introduces the following postulate:

N. {\it Postulate of expansion}: In second approximation, the universe can be described as being made up of ultimate objects whose number increases with time. This formulation takes into account the objections that can be made to postulate J.	

	 It may seem strange that a theory built on principles of very high generality contains a base constant whose value is contingent: the number $ N $ of urs in the universe, with $ N = 10^{120} $. According to the new postulate, $ N $ could now be a measure of the age of the universe. If we measure this age $ t $ by ordinary clocks in nuclear time units, then, presumably, we would have $ N = t^3 $ and if the radius of the world were $ R = t $, we would have a theory where the expansion of the universe is necessarily linked to the creation of matter by the formula $ N = R^3 $, that is to say a cosmology close to those proposed by Dirac and Jordan.

Regarding gravity, one can notice that the empirical link between the curvature of space and the cosmic density of matter is sufficiently close to that required by Einstein's equation to appear more than accidental. As a possible explanation, von Weizs\"{a}cker suggests that it is not the curvature of space that adapts to a given density of matter with a given gravitational constant, but the opposite: in reality, $c$ is this so-called constant which adapts to a given density of matter and a curvature of space. Eddington's famous relation says that the gravitational constant, measured in nuclear units, is $ g = R^{-1} $. We can then expect that: elementary particle physics leads, among others, to a field behaving like the gravitational field (probably containing scalar and tensor components) and that a semantically coherent theory will couple it to the metric tensor. Einstein's gravitational constant was related to the dichotomy between matter and metric field and retained something empirical. One would expect that, in a semantically coherent theory, this constant is reduced to a factor without dimension whose nature is determined by theory.

 In the proposed theory, precisely, gravitation would be linked to the total volume of cosmic space, measured in units, that is to say in number of bits in the world, and its ultimate objects would simply be bits. We would then move on to local fields by considering that matter behaves locally as if it were part of a homogeneous universe, but with a metric determined by the local values of the field. In such a universe, the density of matter would therefore be different from the real cosmic average, and the link between the mass tensor and the metric field would be expressed by the Einstein equation that we would thus find locally.	

\section{The philosophy of alternatives and its problems}

In a 1975 text, taken up in the author's collection of major texts in physics – and not in philosophy –, which is quite significant of the interweaving of the two disciplines in quantum physics, the author begins by recalling that the great task of contemporary physics is to unify the two great physical theories of the twentieth century, namely the theory of space and time (special and general relativity, cosmology) and the theory of different states physical systems and their change over time (quantum mechanics) in a third theory which would allow us to deduce which elementary particles can exist. It is in fact in the theory of elementary particles conceived as quantum field theory that relativity and quantum mechanics must simultaneously apply. At the time when von Weizs\"{a}cker worked on these problems, the theory still only concerned simple fields, the question of interactions not yet being decided. Unification seemed to be able to be envisaged from consideration of symmetry groups.

The relativistic quantum theory being invariant under the Poincaré group, it is also invariant, in the simplest case where we have zero masses, under the conformal group SO(4,2)/$\mathbb{C}^2$, a group which seemed to have gained in importance at the time by Segal's use of it in cosmology and Penrose's considerations in his theory of twistors. On the other hand, the Hilbert space of the simplest possible quantum object – a {\it binary alternative} – admitting the group SU(2) which is, as we have seen, homomorphic to the Euclidean three-dimensional rotation group SO(3), the combination of these two objects gave SU(2,2) or SO(4,2) as symmetry groups.

The question of the unification of relativity and quantum mechanics today receives extremely elaborate mathematical answers – and many more, of course than in the days of von Weizs\"{a}cker. So his attempt is not worth so much by its formalisms as by the epistemological reflection which he introduced. The usual language being the one in which we interpret our physical theories, we must aim for what he calls a "semantic consistency", that is to say, a certain coherence between the mathematical formalisms and the interpretation which one gives of them, this one being especially a goal more than a reality.

To solve the challenge posed by both relativistic and quantum field theory, von Weizs\"{a}cker guessed that it was necessary to better understand quantum physics itself, meaning that he was able to introduce, as we have seen, the idea of a “circumstantial logic” (we prefer this adjective to “temporal” which designates a very particular type of logic, different from that targeted by von Weizs\"{a}cker).

The basis of all quantum axioms being the notion of probability, von Weizs\"{a}cker tried to show first that the latter is only a refinement of the logical modalities of the future in a circumstantial logic. Indeed, the notion of {\it experiment} presupposes a distinction between past and future. In agreement with Aristotle, von Weizs\"{a}cker defines the propositions concerning the future – for example "there will be a naval battle tomorrow" – in terms of modalities of the type "possible", "necessary" or "impossible", which means – Aristotle had already noticed it – that one cannot apply to them the concept of "third-excluded". This is a very different case from the notion of probability, which holds for sets of sets.	
	But quantum mechanics is nothing else than a theory of probabilities. In this sense, in a quantum axiomatics, one does not need to know that there are moments, positions, three-dimensional space, etc. The only thing to know is that there are singular cases and overall statistics, and that all of this is likely to change over time. This consideration of time is important: for von Weizs\"{a}cker, quantum physics is essentially the theory containing the rules which allow us to speak about future events in such a way that we can experimentally verify these statements.

Knowing that, independently of the abstract theory there exist particles which are approximately localized in a 3-dimensional Euclidean space capable of being unified with time in the Minkowski space-time continuum (itself perhaps tangent to a Riemannian manifold), the problem becomes how to deduce these structures from quantum theory. We have already seen that, starting from the fundamental alternatives or urs‚ it was possible, according to von Weizs\"{a}cker, to join the unitary special group SU(2), and then, by homomorphism, the orthogonal special group SO(3) of rotations in a three-dimensional space. What are, however, the problems encountered in such an approach, and which, in particular, remain unresolved? This is the subject of von Weizs\"{a}cker reflection in this 1975 article.
	
	The first thing to consider is that an ur does not identify with a particle, nor with a state for that matter. A particle – for example a free neutrino – exhibits an infinite series of states. But each of them corresponds to a well-defined number of urs and anti-urs, this number differing from state to state. Consequently, the simplest of particles does not correspond to a well-defined number of urs: "the number of urs cannot be defined by the nature of the particle", writes von Weizs\"{a}cker. It is defined – if by chance it is defined – by the particular state of the particle, but in most states it will remain undefined because there will be no eigenvalues for the operator defining the number of urs.

Moreover, as long as one considers particles without mass, one cannot go very far and the objects which one considers are not truly real. The advantage of working with particles of zero mass is that 0 is the only invariant mass value under the conformal group. The introduction of mass particles breaks the conformal symmetry, which means that we then place ourselves in tangent space at particular points where we fix the mass values. This actually means that the concept of particle is a local concept, not a global one.	

	In principle, as we saw above, the total number of urs in the world, according to Von Weizs\"{a}cker is approximately fixed ($10^{120}$) as well as that of nucleons ($10^{80}$) and the radius of the universe ($10^{40}$ Planck lengths). But, we also know that the universe is expanding. We must therefore imagine that new urs are produced at every moment. So we must assume a real production of urs in the universe, the law of interaction not preserving their number. We can call "chronon" the event which produces an additional urs (or Qu-bit), then to make compatible this indefinite production of urs and the invariance characteristic of any quantum theory.

Finally, a last problem arises with the opposition between the mass spectrum, which is continuous (any particle that can take any mass\footnote{This could appear as an analogue of the situation described by Deleuze in his philosophy, when any predicate can be attributed to anything, but, on the one hand, in von Weizs\"{a} cker's quantum theory, it is only about the value of the physical masses and, on the other hand, it is only a temporary situation.} and the empirical observation that there is a small number of discrete masses in the universe. A possible answer is that one should not consider the mass spectrum independently of a theory of the interaction and one could bring into play here statistical considerations analogous to those which exist in thermodynamics. Even so, the problem will not be easy to solve. And it would still be necessary to try to explain the existence of precise masses of stable particles in connection with a cooperative effect of $10^{40}$ or $10^{120}$ urs, which would launch into a kind of thermodynamics of urs.
	
	\section{Beyond von Weizs\"{a}cker: towards a physics of pure information}

Carl Friedrich von Weizs\"{a}cker died in 2007, leaving behind a work which is not only a scientific one. Co-founder of the {\it Society of Germanic Scientists} in 1959, known to have campaigned against the possession of nuclear weapons by Germany, he was also, in the latter part of his life, a philosopher appreciated for his reflection and his ethical positions on various subjects, including international ones. His remarkable personality is perhaps also one of the reasons why his memory continues to be maintained thanks to a foundation, the {\it Foundation Carl Friedrich von Weizs\"{a}cker}, which manages its intellectual heritage, organizes symposia and develops projects on the main challenges of modernity. In addition to the collections of philosophical and physical articles assembled by Michael Drieschner, some works have recently been able to show that the physical theory of which he is the initiator had been better understood in recent years, and developed further than was able to lead von Weizs\"{a}cker himself.

Holger Lyre, for example, briefly summarized von Weizs\"{a}cker's project and its two parts as follows:

First, there is the theory of alternatives. The central idea is that physics is reduced to predicting the results of measurements. The results of the measurements can be reformulated in terms of alternatives with $ n $ empirically decidable components, each of which can be reduced to a (Cartesian) product of binary alternatives. Binary alternatives, in turn, can be thought of as bits of information (Qu-bits when it comes to quantum alternatives). Thus physics is reduced to information or, more precisely, to potential information (see \cite{Lyr}, 2).

Then there is the idea that the quantum theory of binary alternatives is the theory of objects in three-dimensional space. This second point has already been briefly explained in the commentary on Heisenberg's text from which we started. But we can still formulate it here in a more rigorous way by noting (see \cite{Lyr1}) that a fundamental element (ur) is described by a state vector of a complex 2-dimensional space of the type:

	\begin{equation}
| u_{r}\rangle \in \mathbb{C}^2, \qquad	r = 1, 2.
	\end{equation}
	Any state can then be represented by means of a Hilbert space which is a subspace of the space of the tensor product of 2-dimensional Hilbert spaces belonging to urs. So we have:
\begin{equation}
V^m \subseteq T^n = \otimes_{n} \mathbb{C}^2,	\qquad m \leq 2^n.
\end{equation}

The symmetry group $ Q $ of an ur $ u_{r} $ leaves the unit form invariant:

\begin{equation}
\langle u|u\rangle = u_{1}^*u_{1}+ u_{2}^*u_{2},
\end{equation}
and contain the subgroups:
\begin{equation}
\textnormal{SU(2)} \times \textnormal{ U(1) \   and } \ K.	
\end{equation}

The anti-linear transformations \^{K} $ \in K $ then act like \^{K} $ | u \rangle = i \hat {\sigma}_{2} $ | $ u^* \rangle $, where $ \hat {\sigma} _ {2}$ is the second Pauli matrix and * the complex conjunction.

In this theory, the three-dimensional space where the objects are located is derived as a consequence of these mathematical conditions. This can be explained by analyzing the concept of space. In most cases, the spatial distance between objects can be understood as the parameter of the interaction they have with each other. On the other hand, the definition of a physical object (for example, a massive elementary particle) depends on the spatial range it occupies.

 Assuming that all objects consist of urs, the total state of the universe should remain unchanged by transforming all urs with the same operator of the ur symmetry group, which is essentially SU(2). Thus, the interaction between all objects must be invariant and therefore the position space as a parameter space for the interaction force must have the same structure as the symmetric space of the symmetry group of urs. In von Weizs\"{a}cker theory, one therefore supposes that the position space is identified with the homogeneous space $\mathbb{S}^3$ of the group SU(2), and temporal development of urs will be described by the group of phase transformations U(1). Knowing that:
 
 \[
\textnormal{SU}(2) \times \textnormal{U}(1) = \mathbb{S}^3 \times \mathbb{S}^1.
\]
	
	SO(3) giving access to 3-dimensional space, U(1) should lead to temporal considerations. But, as Holger Lyre points out, it is relatively implausible that U(1) could be a model of cosmic time which, then, would be cyclical.

	Starting again from the equality SU(2) = $ \mathbb {S}^3$ as a model of the global space, we can then understand the urs as nonlocal functions on SU(2) represented by dyads of spinors (composed of spinors $u^A, v^A$ satisfying $u_ {A} v^A = -v_ {A} u^A = 1$). Lyre then points out that a dyad of spinors is equivalent to a tetrad of zero vectors, where the four zero vectors have the spinor form, but consist of mixed combinations of $ u_{A}$ and $v_ {A}$. Considering appropriate linear combinations of zero vectors, such a zero tetrad can generally be written in a real-valued form $\theta_{\mu}^{\alpha} = (\theta _{\mu}, x_{\mu} , y_{\mu}, z_{\mu})$, where the space vectors $x_{\mu}, y_{\mu}, z_{\mu}$ represent a spatial triplet tangent to $\mathbb{ S}^3$ with a temporal orthogonal vector $\theta_{\mu}$. The interesting point is that, since the tetrad is written in terms of ur-spinor components, a (first) quantification of the urs also induces a quantification of the tetrad. Such a quantized ur-tetrad, however, means nothing more than quantized coordinates in the chosen spacetime model.

	An effect of quantization by ur-tetrad is that the time operator $\hat {t}_{\mu}$ is simply the numeric operator $\hat {n} = (\frac {1} {2} \sum_{r} \hat {a}_{r}^+, \hat {a}_{r})$ in the Fock space of the Bosonian urs. This is consistent with von Weizs\"{a}cker's hypothesis according to which the growth of the total number of urs is a measure of temporal cosmic evolution. Now, since the numerical operator has indeed zero as a lower bound, the global space-time model $\mathbb{S}^3 \times \mathbb{R}^+$ with a time parameter variety $\mathbb{R}^+$ thus seems justified (thereby avoiding globally closed time-type curves as mentioned above) (see \cite{Lyr}).

	It is, in spite of everything, difficult, from there, to describe the (quantum)  gravity  in the theory of urs. One could of course think that $\theta_{\mu}^{\alpha}$ could represent four vector bosons, that is to say {\it gravitons} without mass of spin 1, and thus obtain a corresponding wave equation $\theta_{\mu}^{\alpha} (x) = 0$ analogous, for example, to that of Klein-Gordon for a massless universe. Perhaps this could describe the gravitational field in a nonstandard form (i.e. not as a spin-2 field) in its linearized boundary. However, what is really sought is rather a recipe that allows this field to be made dynamic and to couple it to matter. Without going into details here (see \cite{Lyr}), the thing is surely much more difficult because it would require a gravity gauge theory.

	According to Holger Lyre, a good place to start could be the fact that the quantized ur-tetrad generates a group that could be used as a gauge group. Indeed, the Lie algebra of the ur-tetrad operators is 12-dimensional and the corresponding Lie group is isomorphic to $\textnormal {SL} (2, \mathbb {C}) \times \textnormal{SL} (2, \mathbb {C}$). Whether and how this group can be used for an appropriate gauging approach, however, remains an open question. The operators of this algebra acting in the flat Minkowski space-time, this results in ur-theory by unimodular groups, rather than unitary ones. The gravitational field being thus described in a flat space, this also seems to mean that it is perhaps a local field, just like the other Yang-Mills fields.	

	We will also note as a last problem that in ur-theory, as soon as we start from $\mathbb{S}^3$, it is in principle impossible to arrive at a global spherical model of the cosmos \footnote {Mathematically , we can always project $\mathbb {S}^3$ onto $\mathbb{S}^2$ (Hopf fibration), but, physically, we do not see what could justify it.}.

	It remains to be seen how this ur-theory of von Weizs\"{a}cker leads to a theory of pure information and an informational universe.

	The links between space and information are now introduced through the question of the entropy of black holes. Initially, black holes were only characterized by three quantities: mass, angular momentum, and charge. But, as Jacob Bekenstein noted in 1973, there are a number of similarities between black hole physics and thermodynamics, the most striking being the kinship of black hole zone behaviors and entropy, the two quantities tending to increase irreversibly in the course of time (see \cite {Bek}). The area of the event horizon, denoted $ A $, expressed in Planck units, has thus been found to be an appropriate measure of the entropy content of the black hole, giving rise to the very simple formula:	
	
	\begin{equation}
S = \frac{1}{4} A,
\end{equation}

leading at the same time to a second generalized law of thermodynamics. As there is a total parallelism between entropy and information (see \cite{Par2}, this formula amounts to characterizing a physical object (in this case, a black hole) in terms of information.

Suppose then that the universe is some kind of black hole. One can then calculate its informational content in Planck units as follows. If we assume a spherical symmetry, hence a surface $ A = 4 \pi R^2 $ and a Schwarzchild radius $ R = 2M $, Beckenstein's formula becomes:

\begin{equation}
S_{u} = \frac {1} {4} 4 \pi (2m)^2 = 4 \pi m^2.
\end {equation}

With, for the whole universe, $ M_{u} = 10^{60} m_{0} $, we finally get $ S_{u} = 10^{120}$ bits, according to what von Weizs\"{a}cker had, by another way, already obtained in the years 1960 and which we recalled above.

It seems, in the end, that formula (8) above poses an equivalence between matter-energy, on the one hand, and information, on the other hand. We cannot however, in our opinion, see in it anything other than two different ways of reading the universe: information is not, strictly speaking, matter-energy. Unless, of course, one sees in the urs or fundamental alternatives of von Weizs\"{a}cker, not simple theoretical choices but concrete bifurcations, real modes of embodied beings. But to go further, we must then get out of the purely transcendental point of view and rediscover that of ontology.

\end{document}